\documentclass{ws-ijqi}
\usepackage{graphicx}
\usepackage{epsfig}

\begin{document}
\newcommand{\beq}{\begin{equation}}
\newcommand{\eeq}{\end{equation}}

\markboth{Juha Vartiainen}{Acceleration of quantum algorithms using three-qubit gates}

\catchline{}{}{}{}{}
\title{ACCELERATION OF QUANTUM ALGORITHMS USING THREE-QUBIT GATES}
\author{Juha J. Vartiainen}
\address{Materials Physics Laboratory, Helsinki University of Technology, \\
POB 2200 (Technical Physics), FIN-02015 HUT,
Finland\\ juhav@focus.hut.fi}
\author{Antti\ O.\ Niskanen}
\address{VTT Information Technology, Microsensing, POB 1207,
02044 VTT, Finland}
\author{Mikio Nakahara}
\address{Department of Physics, Kinki University,
Higashi-Osaka 577-8502, Japan}
\author{Martti\ M.\ Salomaa}
\address{Materials Physics Laboratory, Helsinki University of Technology, \\
POB 2200 (Technical Physics), FIN-02015 HUT,
Finland}

\maketitle

\begin{abstract}
Quantum-circuit optimization is essential for any practical realization of
quantum computation, in order to beat decoherence. We present a scheme for implementing the final stage
in the compilation of quantum circuits, i.e., for finding the actual physical
realizations of the individual modules in the quantum-gate library. We find that numerical optimization can be
efficiently utilized in order to generate the appropriate control-parameter sequences which produce the
desired three-qubit modules within the Josephson charge-qubit model. Our work suggests ways in
which one can in fact considerably reduce the number of gates required to implement a
given quantum circuit, hence diminishing idle time and significantly accelerating the execution
of quantum algorithms.
\end{abstract}
\keywords{decoherence, Josephson charge qubit, multiqubit quantum gates, numerical optimization}

\section{Introduction}

The most celebrated and potentially useful quantum algorithms,
which include Shor's factorization algorithm\cite{Shor} and
Grover's search\cite{Grover}, manifest the potential of a quantum
computer compared to its classical counterparts.

Widely different physical systems have been proposed to be
utilized as a quantum computer\cite{Clark,Martini}. The main
drawback shared by most of the physical realizations is the short
decoherence time. Decoherence\cite{zurek} destroys the pure
quantum state which is needed for the computation and, therefore,
strongly limits the available execution time for quantum
algorithms. This, combined with the current restricted technical
possibilities to construct and control nanoscale structures,
delays the utilization of quantum computation for reasonably
extensive\cite{DiVincenzo} algorithms.

The execution time of a quantum algorithm can be reduced by
optimization. The methods similar to those common in classical
computation\cite{Aho} can be utilized in quantum compiling,
constructing a quantum circuit\cite{Deutsch} for the algorithm.
Moreover, the physical implementation of each gate can and must be
optimized in order to achieve gate sequences long enough, for
example, to implement Shor's algorithm within typical decoherence
times\cite{PRL}.

Any quantum gate can be implemented by finding an elementary gate
sequence~\cite{elementary,Zhang} which, in principle, exactly
mimics the gate operation. In the most general case on the order
of $4^n$ elementary gates are needed to implement an arbitrary
$n$-qubit~\cite{Shende}. Fortunately, remarkably shorter
polynomial gate sequences are known to implement many commonly
used gates, such as the $n$-qubit quantum Fourier transform (QFT).
In addition to the exact methods, quantum gates can be implemented
using techniques which are approximative by
nature~\cite{PRL,Harrow,Knill,Burkard}.

In this paper we consider the physical implementation of
nontrivial three-gate operations. As an example of the power of
the technique, we show how to find realizations for the Fredkin,
Toffoli, and QFT gates through numerical optimization. These gates
have been suggested to be utilized as basic building blocks for
quantum circuits and would thus act as basic extensions of the
standard universal set of elementary gates. However, the method
presented can be employed to find the realization of any
three-qubit gate. Having more computer resources available would
allow one to construct gates acting on more than three qubits.

The numerical method allows us a straightforward and efficient way
for finding the physical implementation of any quantum gate. Thus,
the method may prove to be practical or even necessary for an
efficient experimental realization of a quantum computer.

We concentrate on a hypothetical Josephson charge qubit
register\cite{schon}, since the experimental investigations of
superconducting qubits is active; see, for instance, Refs.
\refcite{Pashkin} \-- \refcite{martinis}. The scheme utilizes the
number degree of freedom of the Cooper pairs in a superconducting
Josephson-junction circuit. It is potentially scalable and it
offers, in principle, full control over the quantum register.
Moreover, the method employed here is easily extended to any
physical realization providing time-dependent control over the
physical parameters.

\section{\label{Sec:Phys} Physical Model}

The physical implementation of
a practical quantum algorithm requires that it is decomposed into modules whose physical
realizations are explicitly known.
In the quantum computer, the gate operations are realized through
unitary operations $U$ that result from the temporal evolution of the
physical state of the quantum register. The unitary evolution is
governed by the Hamiltonian matrix, $H(\gamma)$, which describes the energy of the
system for a given setting of physical parameters $\gamma$. In general, the
parameters are time-dependent, $\gamma=\gamma(t)$. The induced unitary operator is
obtained from the formal solution of the Schr\"odinger  equation
\beq
\label{eq:U}
U=\mathcal{T}
\exp\left(-i\int_{\gamma(t)} H(\gamma(t))dt\right),
\end{equation}
where $\mathcal{T}$ stands for the time-ordering operator and we have chosen
$\hbar=1$.

\begin{figure}[th]
\centerline{\psfig{file=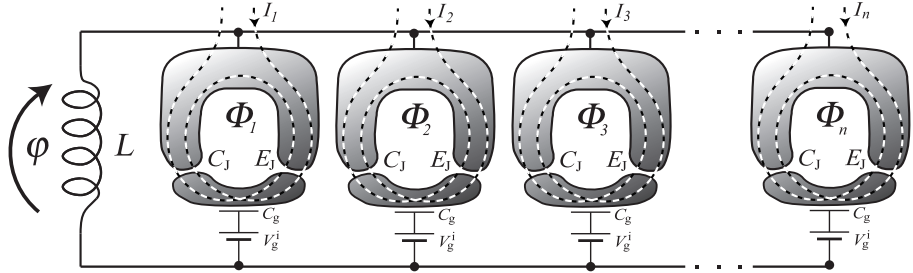,width=0.95\textwidth}}
\vspace*{8pt} \caption{\label{fg:kupitit} Schematic of an array of
Josephson charge qubits coupled in parallel with an inductor.}
\end{figure}

We consider the Josephson charge qubit register as a realization
of a quantum computer, see Fig.~\ref{fg:kupitit}. The register is
a homogenous array of mesoscopic superconducting islands and the
states of the qubit correspond to either zero or one extra Cooper
pair residing on the island. Each of the islands is capacitively
coupled to an adjustable gate voltage, $V_{\rm g}^i(t)$. In
addition, they are coupled to a superconducting lead through
mesoscopic SQUIDs. We consider an ideal situation,
where each Josephson junction in the SQUID devices has
the same Josephson energy $E_{\rm J}$ and capacitance $C_{\rm J}$.
The magnetic flux $\Phi_i(t)$
through the $i^{\rm th}$ SQUID loop is a control parameter which may be produced
by adjustable current $I_i$. The qubit array is coupled in parallel with an inductor, $L$,
which allows the interaction between the qubits.

In this scheme the Hamiltonian for the qubit register is\cite{schon,PRL}
\begin{equation}
H=\sum_{i}^n   \Bigl \lbrace -\frac{1}{2}B^i_z\sigma_z^i-\frac{1}{2}B^i_x\sigma_x^i \Bigr \rbrace -
\sum_{i \neq j}^{n,n} CB^i_xB^j_x\sigma_y^i\otimes \sigma_y^j ,
\label{eq:ham}
\end{equation}
where the standard notation for Pauli matrices has been utilized
and $\sigma_x^i$ stands for $I \otimes \ldots \otimes \sigma_x
\otimes I \ldots \otimes I$. Above, $B^i_x$ can be controlled with
the help of a flux $\Phi_i(t)$ through the $i$th SQUID, $B^i_z$ is
a tunable parameter which depends on the gate voltage $V_{\rm
g}(t)$ and $C$ is a constant parameter describing the strength of
the coupling. We set $C$ equal to unity by rescaling the
Hamiltonian and time. The approach taken is to deal with the
parameters $B^i_z$ and $B^i_x$ as dimensionless control
parameters.

In the above Hamiltonian, each control parameter can be set to zero, to the degeneracy point,
thereby eliminating all temporal evolution. The implementation of one-qubit operations is straightforward through the
Baker-Campbell-Hausdorff formula, since the turning on of the parameters $B^i_z$ and $B^i_x$ one by one does not
interfere with the states of the other qubits.
Implementation of two-qubit operations is more complex since simultaneous application of nonzero parameter values for many qubits causes
undesired interqubit couplings. However, by properly tuning the parameters it is possible to compensate the interference and to perform
any temporal evolution in this model setup. This is partly why numerical methods are necessary for finding the required control-parameter sequences.

Finally, we point out that using the above Hamiltonian we are able to perform gates $U \in SU(2^k)$ since
the Hamiltonian is traceless. However, for every gate $U \in U(2^k)$ we can find a matrix $U'=e^{i\phi}U$ which has
a unit determinant. The global phase factor $e^{i\phi}$ corresponds to redefining the zero level of energy.

\section{Numerical Methods}

\begin{figure}[th]
\centerline{\psfig{file=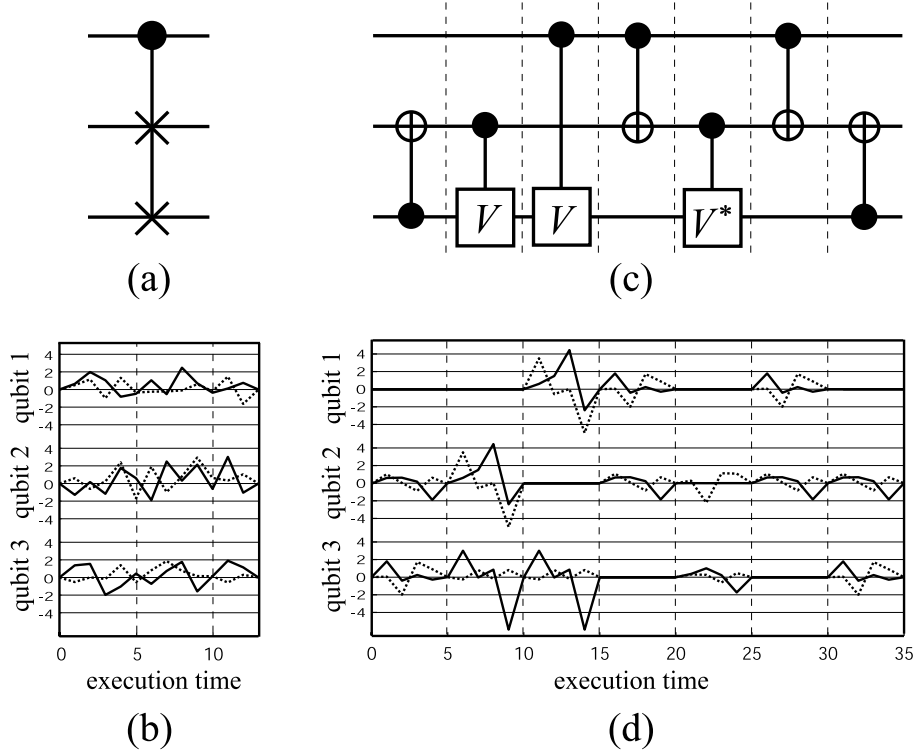,width=0.95\textwidth}}
\vspace*{8pt} \caption{\label{fg:touhotus} Implementation of the
Fredkin gate on the Josephson charge qubit model. (a) The quantum
circuit symbol of the Fredkin gate, and (b) its physical
implementation by controlling all three qubits simultaneously. (c)
The two-qubit gate decomposition of the Fredkin gate. Here
$V=\sqrt{\sigma_x}$ and $V^*$ stands for its Hermitian conjugate.
(d) The physical implementation of the gate sequence; note that
during each gate operation, one of the qubits is in the idle
state. The vertical axis in figures (b) and (d) stands for the
control parameter field amplitudes; the solid line describes the
parameter $B_z^i$ and the dotted line the parameter $B_x^i$, see
text.}
\end{figure}

We want to determine the physical realization for the quantum gates.
Our aim is to numerically solve the inverse problem of finding the
parameter sequences $\gamma(t)$ which would yield the desired
gate operation when substituted into Eq.~(\ref{eq:U}). The numerical
optimization provides us with the realizations for not only any one- and
two-qubit, but also for any three-qubit gates. Using the three-qubit
implementation we circumvent the idle time in qubit control which
provides us faster execution times, see Fig.~\ref{fg:touhotus}.

In the Josephson charge qubit model the Hamiltonian for the
$n$-qubit register, Eq.~(\ref{eq:ham}), depends on the external
parameters $\gamma(t) = [B^1_z(t) \ldots B^n_z(t); \, B^1_xz(t)
\ldots B^n_x(t)]$. To discretize the integration path $\gamma(t)$
for numerical optimization we consider a parametrization in which
the values of the control-parameter fields, $\{ B^i_z(t) \}$ and
$\{ B^i_z(t) \}$, are piecewise linear functions of time.
Consequently, the path $\gamma(t)$ can be fully described by a set
of parameter values at $\nu$ control points, where the slopes of
the fields changes. We denote the set of these values collectively
as $X_\gamma$. To obtain a general $k$-qubit gate $U_k \in
SU(2^k)$ one needs to have enough control parameters to
parameterize the unitary group $SU(2^k)$, which has a total of
$2^{2k}-1$ generators. Since there are $2k$ free parameters for
each control point in $\gamma$ we must have \beq \label{condition}
2k\nu\geq 2^{2k}-1. \eeq We use $\nu=12$ for the three-qubit gates
and $\nu=4$ for the two-qubit gates. We force the parameter path
to be a loop, which starts from and ends at the degeneracy point,
where all parameter values vanish. Then we can assemble the
modules in arbitrary order without introducing mismatch in the
control parameters. We further set the time spent in traversing
each interval of the control points to equal unity. Eventually,
the execution time of $U_k$ is proportional to $\nu+1$, which
gives us a measure to compare different implementations.
Figure~\ref{fg:touhotus} illustrates our approach and shows the
benefits of the three-qubit implementation of the Fredkin gate
compared to corresponding implementation through two-qubit gate
decomposition. Note that the two-qubit gate implementation could
be further optimized\cite{Smolin}.

We evaluate the unitary operator in Eq.~(\ref{eq:U}) in a numerically robust
manner by dividing the loop $\gamma(t)$ into tiny intervals that
take time $\Delta t$ to traverse. If $\gamma_i$ denotes all the
values of the parameters in the midpoint of the $i^{th}$ interval,
and $m$ is the number of such intervals, we then find to a good
approximation
\beq
U_{X_\gamma}\approx \exp(-iH(\gamma_m)\Delta t)\ldots \exp(-iH(\gamma_1)\Delta t).
\end{equation}
The evaluation of the $U_{X_\gamma}$ consists of independent matrix multiplications which
can be evaluated simultaneously. This allows straightforward parallelization of the computation.
To calculate the matrix exponentials efficiently we use the truncated Taylor-series expansion
\beq
e^A  \approx \sum_{k=0}^m \frac{A^k}{k!},
\label{eq:exptay}
\end{equation}
where $m$ is an integer in the range $3$ -- $6$. Since the
eigenvalues of the anti-Hermitian matrix $A=-iH\delta t$ are
significantly less than unity, the expansion converges rapidly.
The applicability of the approximation can be confirmed by
comparing the results with the exact results obtained using
spectral decomposition.

\begin{figure}[th]
\centerline{\psfig{file=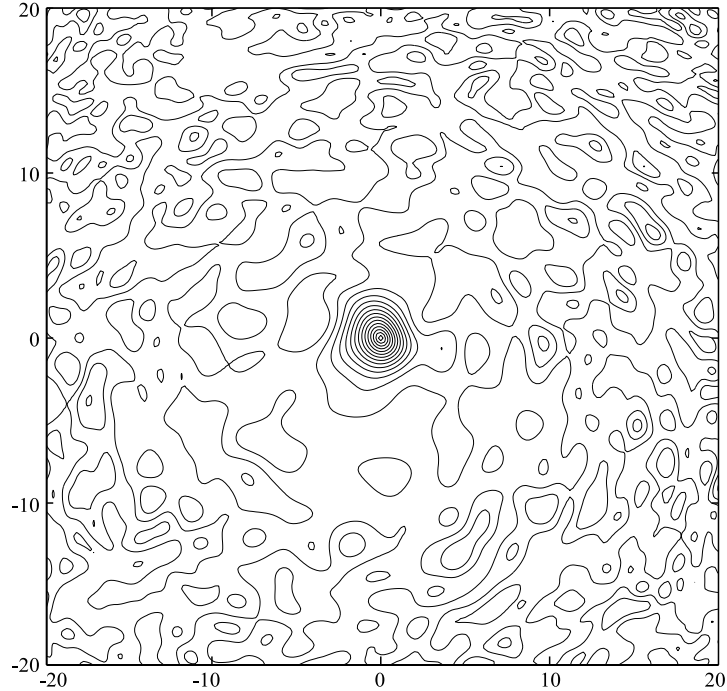,width=0.75\textwidth}}
\vspace*{8pt}
\caption{\label{fg:ersur} Typical planar cut of the error function
space. The plane through the minimum point $X_{\rm min}$ has been
chosen arbitrarily in the parameter space. The irregular shape of
the landscape easily reveals the complexity of finding the global
minimum and the reason why the gradient-based methods fail.}
\end{figure}

Using the above numerical methods we transform the inverse problem of finding the
desired unitary operator into an optimization task. Namely, any $\hat{U}$ can be found as the
solution of the problem of minimizing the error function
\beq
\label{eq:error} f(X_\gamma)=\|\hat{U}-U_{X_\gamma}\|_F
\eeq over
all possible values of $X_\gamma$. Here $\|\cdot\|_F$ is the
Frobenius trace norm defined as
$\|{A}\|_F=\sqrt{\mathrm{Tr}\left({A}^\dagger{A}\right)}$.
The minimization landscape is rough, see Fig.~\ref{fg:ersur}. Thus we apply the robust
polytope search algorithm\cite{poly} for the minimization. We have assumed that a suitable limit
of sufficient accuracy for the gate operations is given by the requirement of the applicability\cite{DiVincenzo} of
quantum error correction
\beq
\|U_{X_\gamma}-\hat{U} \|<10^{-4},
\eeq
where $\hat{U}$ and $U_{X_\gamma}$ are the target and the numerically optimized gate operations, respectively.

\section{\label{sec:results}Quantum Gate Optimization Results}

We have applied the minimization procedure to various three-qubit
gates and found that the error functional of Eq.~(\ref{eq:error}) can be minimized
to values below $10^{-4}$ by running the polytope search repetitively.
Table~\ref{tbl:fredkinarvot} represents the optimized control parameters which serve to yield the Fredkin gate
when applied to the Josephson charge qubit Hamiltonian. Numerical results for the Toffoli and three-qubit
QFT gates are represented in Tables~\ref{tbl:toffoliarvot} and~\ref{tbl:QFTarvot}, respectively.
Finding the control parameter using the polytope search requires on the order
of $10^6$ error-function evaluations, which takes tens of hours of CPU time, but can be done in
a reasonable time by using parallel computing.

We found that the error functional grows linearly in the vicinity
of the minimum point $X_{\gamma}$, which implies that the parameter sequence found
may be robust. The robustness was further analyzed by adding
Gaussian noise to the control parameters of the path $\gamma(t)$. Such a
sensitivity analysis confirmed that the error scales linearly with
the root-mean-square amplitude of the surplus Gaussian noise.

\begin{table}[h]
\tbl{\label{tbl:fredkinarvot} Field amplitudes at the control points for the Fredkin gate.}
{\begin{tabular}{@{}ccccccc@{}}
\toprule
time &$ B_z^1 $&$ B_z^2 $&$ B_z^3 $&$ B_x^1 $&$ B_x^2 $&$ B_x^3 $\\
\colrule
1 & 0.00000 & 0.00000 & 0.00000 & 0.00000 & 0.00000 & 0.00000 \\
2 & 0.71637 & -1.44846 & 1.54511 & 0.55428 & 0.67228 & -0.58105 \\
3 & 2.23337 & 0.18377 & 1.73522 & 1.29275 & -0.69463 & 0.01513 \\
4 & 1.17895 & -1.31725 & -2.22145 & -1.11461 & 0.27210 & -0.18665 \\
5 & -0.92555 & 1.97326 & -1.15875 & 1.49438 & 2.69507 & 1.57872 \\
6 & -0.54804 & 0.66834 & 0.48872 & -0.38981 & -1.88659 & -0.60226 \\
7 & 1.18034 & -2.13101 & -0.81205 & -0.27817 & 2.13894 & 0.92208 \\
8 & -0.59994 & 2.80989 & 0.82839 & -0.24260 & -1.09419 & 2.09561 \\
9 & 2.78429 & 0.35914 & 1.98896 & -0.11839 & 0.90439 & 0.83671 \\
10 & 0.79364 & 2.40575 & -1.78131 & 0.67600 & 3.31481 & 0.17828 \\
11 & -0.41098 & -0.69585 & 0.15594 & -0.21996 & 0.70917 & 0.15377 \\
12 & 0.12630 & 3.39809 & 2.14043 & 1.65229 & 0.37794 & -0.64223 \\
13 & 0.84941 & -1.17701 & 1.28801 & -1.84075 & 1.16739 & 0.33965 \\
14 & 0.00000 & 0.00000 & 0.00000 & 0.00000 & 0.00000 & 0.00000 \\
\botrule
\end{tabular}}
\end{table}

\begin{table}[h]
\tbl{\label{tbl:toffoliarvot} Field amplitudes at the control points for the Toffoli gate.}
{\begin{tabular}{@{}ccccccc@{}} \toprule
time &$ B_z^1 $&$ B_z^2 $&$ B_z^3 $&$ B_x^1 $&$ B_x^2 $&$ B_x^3 $\\
\colrule
1 & 0.00000 & 0.00000 & 0.00000 & 0.00000 & 0.00000 & 0.00000 \\
2 & 0.00286 & -0.06484 & 0.96050 & 0.72386 & 0.33310 & -0.22026 \\
3 & 2.85647 & -0.08874 & 2.94358 & 1.60795 & -0.18192 & 0.03931 \\
4 & 0.67879 & -1.70364 & -2.54280 & -1.65771 & -0.04722 & -0.25411 \\
5 & -0.17379 & 0.87916 & 0.19581 & 1.55484 & 2.98447 & 1.22991 \\
6 & 0.01847 & 2.68973 & -0.18098 & 0.02898 & -0.54301 & -0.15977 \\
7 & 0.21569 & -3.27483 & -0.33407 & -0.31173 & 2.26503 & 0.32031 \\
8 & -0.57439 & 4.25644 & 1.25986 & 0.12262 & 0.06238 & 1.87619 \\
9 & 3.40836 & -0.48759 & 0.44296 & -0.20867 & 0.04664 & 1.00381 \\
10 & -0.60520 & 1.59369 & 0.87620 & 0.95412 & 2.75968 & 0.37209 \\
11 & -0.10762 & 0.16258 & -0.24672 & -0.11839 & 1.38245 & 0.01990 \\
12 & 0.20275 & 1.97553 & 1.12769 & 1.07003 & 0.46081 & -0.35437 \\
13 & 0.99088 & -0.23145 & 0.68050 & -2.12999 & 0.74237 & 0.01537 \\
14 & 0.00000 & 0.00000 & 0.00000 & 0.00000 & 0.00000 & 0.00000 \\
\botrule
\end{tabular}}
\end{table}

\begin{table}[h]
\tbl{\label{tbl:QFTarvot} Field amplitudes for the three-qubit QFT gate.}
{\begin{tabular}{@{}ccccccc@{}} \toprule
time &$ B_z^1 $&$ B_z^2 $&$ B_z^3 $&$ B_x^1 $&$ B_x^2 $&$ B_x^3 $\\
\colrule
1 & 0.00000 & 0.00000 & 0.00000 & 0.00000 & 0.00000 & 0.00000 \\
2 & 0.49824 & 0.41039 & 1.75837 & 0.42339 & 0.67345 & 1.83257 \\
3 & -0.18007 & 0.55372 & -1.79297 & 0.64987 & 0.53048 & -0.39300 \\
4 & 0.73625 & 0.60488 & -0.94171 & 0.61458 & 0.09641 & -0.39863 \\
5 & 2.21744 & 1.28419 & 2.82723 & 0.47046 & 1.04206 & 1.59345 \\
6 & 0.47037 & -0.48092 & -0.53215 & 0.04297 & 0.21802 & 1.24063 \\
7 & 0.69085 & 0.72558 & 1.00427 & 0.22332 & 1.25082 & -0.25144 \\
8 & 2.61154 & 0.87134 & 0.74335 & 0.31834 & -0.00374 & 1.64643 \\
9 & 0.24827 & 0.82952 & 1.04102 & 2.31043 & 1.00804 & 0.98377 \\
10 & -0.90785 & -1.32491 & 1.10923 & 0.69935 & -0.15359 & -0.34420 \\
11 & 0.59315 & 1.36082 & -0.19764 & 1.83023 & 0.58541 & 0.85453 \\
12 & 0.76819 & 0.31529 & 0.24531 & -0.40221 & 1.13052 & 0.68184 \\
13 & -0.85651 & 0.02093 & 0.85491 & 1.33447 & 0.56580 & 0.06332 \\
14 & 0.00000 & 0.00000 & 0.00000 & 0.00000 & 0.00000 & 0.00000 \\
\botrule
\end{tabular}}
\end{table}

In our scheme, any three-qubit gate requires an integration path
$\gamma(t)$ with 12 control points, which takes 13 units of time
to execute. Similarly, a two-qubit gate takes 5 units of time to
execute. Table~\ref{tbl:compare} summarizes our results by
comparing the number of steps that are required to carry out a
single three-qubit gate or using a sequence of two-qubit gates.
The results are calculated for the Fredkin and Toffoli gates
following the decomposition given in Refs.~\refcite{Smolin} and
\refcite{elementary}. For a QFT gate the quantum circuit is
explicitly shown, for example, in Ref.~\refcite{Gruska}. Any
three-qubit gate can be realized by using 68 controlled$^2$$U$ and
controlled$^2$ NOT gates. This number can be reduced to 50 using
palindromic optimization\cite{Svore}. The decomposition of the
controlled$^2$$U$ gate is discussed in Ref.~\refcite{elementary}.
Note that the results in Table~\ref{tbl:compare} are calculated
assuming that the physical realization for any two qubit modules
is available through some scheme similar to the one which is
employed in this paper and one-qubit gates are merged into
two-qubit modules. The implementation of a general two-qubit
module using a limited set of gates, for example, one-qubit
rotations $R_y$ and $R_z$ and the CNOT gate has recently been
discussed in Ref.~\refcite{Shende}.

\begin{table}[h]
\tbl{\label{tbl:compare} Comparison of the execution times for
various quantum gates. } {\begin{tabular}{@{}cccccc@{}} \toprule
gate & Fredkin & Toffoli & QFT & $U \in SU(2^3)$ & $U \in SU(2^3)$ \\
  & & & & decomposed & 3-qubit gates \\
\colrule
number of two-qubit gates& 5 & 3 & 3 & 206 & -\\
execution time & 25 & 15 & 15 & 1030 & 13 \\
\botrule
\end{tabular}}
\end{table}

\section{Discussion}

We have shown how to obtain approximative control-parameter
sequences for a Josephson charge-qubit register with the help of a
numerical optimization scheme. The scheme utilizes well known
theoretical methods and the results are obtained through heavy
computation. Our method can prove useful for experimental
realization of working quantum computers. The possibility to
implement nontrivial multiqubit gates in an efficient way may well
turn out to be a crucial improvement in making quantum computing
realizable. For example, Josephson-junction qubits suffer from a
short decoherence time, in spite of their potential scalability,
and therefore the runtime of the algorithm must be minimized using
all the possible ingenuity imaginable.

Here we have utilized piecewise linear parameter paths. This makes
the scheme experimentally more viable than the pulse-gate
solutions, since the parameters are adjusted such that no fields
are switched instantaneously. However, the numerical method
proposed for solving the time evolution operator is not unique.
Some implicit methods for the integration in time may turn out to
yield the results more accurately in the same computational time.
Furthermore, for practical applications it may turn out to be
useful to try and describe the parameter paths using a collection
of smooth functions and to find whether they would produce the
required gates.

To summarize the results of our numerical optimization, we
emphasize that more efficient implementations for quantum
algorithms can be found using numerically optimized three-qubit
gates. In the construction of large-scale quantum algorithms even
larger multiqubit modules may prove powerful. The general idea is
to use classical computation to minimize quantum computation time,
aiming below the decoherence limit.

\section*{Acknowledgements}

JJV  thanks the Foundation of Technology (TES, Finland) for a
scholarship and the Emil Aaltonen Foundation for a travel grant to
attend EQIS'03 in Japan; MN is grateful for partial support of a
Grant-in-Aid from the Ministry of Education, Culture, Sports,
Science, and Technology, Japan (Project Nos. 14540346 and
13135215). This research has been supported in the Materials
Physics Laboratory at HUT by the Academy of Finland through the
Research Grants in Theoretical Materials Physics (No. 201710) and
in Quantum Computation (No. 206457). We also want to thank CSC -
Scientific Computing Ltd (Finland) for parallel computing
resources.


\begin{thebibliography}{99}

\bibitem{Shor}
P.~W. Shor, in
{\it Algorithms for quantum computation: Discrete logarithms and factoring --
Proc. 35nd Annual Symposium on Foundations of Computer Science}, ed. S. Goldwasser,
(IEEE Computer Society Press, 1994), p. 124.

\bibitem{Grover}
L.~K. Grover,
{\it Phys. Rev. Lett.} {\bf 79}, 325 (1997).

\bibitem{Clark} R. Clark (Ed.), {\it Experimental Implementation
of Quantum Computation (IQC~'01)}, (Rinton Press Inc., New Jersey,
2001).

\bibitem{Martini} F. De Martini (Ed.), {\it Experimental Quantum
Computation and Information} (International School of Physics
"Enrico Fermi", vol. {\bf 148}) (IOS Press, Amsterdam, 2002).

\bibitem{zurek}
W.~H. Zurek,
{\it Rev. Mod. Phys.} {\bf 75}, 715 (2003).


\bibitem{DiVincenzo}
D.~P. DiVincenzo,
{\it Fortschr. Phys.} {\bf 48}, 771 (2000).


\bibitem{Aho}
A.~V. Aho, R. Sethi, and J.~D. Ullman,
{\it Compilers: Principles, Techniques and Tools},
(Addison-Wesley, Reading, Massachusetts, 1986).

\bibitem{Deutsch}
D.~Deutsch,
\newblock  Proc. R. Soc. London A {\bf 425}, 73 (1989).

\bibitem{PRL}
A.~O. Niskanen, J.~J. Vartiainen, and M.~M. Salomaa,
{\it Phys. Rev. Lett.} {\bf  90}, 197901 (2003);
J.~J. Vartiainen, A.~O. Niskanen, M. Nakahara, and M.~M. Salomaa,
``Implementing Shor's algorithm on Josephson Charge Qubits", quant-ph/0308171.

\bibitem{elementary}
A. Barenco, C. H. Bennett, R. Cleve, D. P. DiVincenzo, N. Margolus, P. Shor, T. Sleator, J. Smolin, and H. Weinfurter,
{\it Phys. Rev. A} {\bf 52}, 3457 (1995).


\bibitem{Zhang}
J. Zhang, J. Vala, S. Sastry, and B. Whaley,
{\it Phys. Rev. Lett.} {\bf 91}, 027903 (2003).

\bibitem{Shende}
V.~V. Shende, I. L. Markov, and S. S. Bullock, ``Minimal Universal
Two-Qubit Quantum Circuits", quant-ph/0308033.

\bibitem{Harrow}
A.~W. Harrow, B. Recht, and I.~L. Chuang,
{\it J. Math. Phys.} {\bf 43}, 4445 (2002).

\bibitem{Knill}
E.~Knill, ``Approximation by quantum circuits", quant-ph/9508006.

\bibitem{Burkard}
G. Burkard, D. Loss, D. P. DiVincenzo, and J. A. Smolin,
{\it Phys. Rev. B} {\bf 60}, 11404 (1999).

\bibitem{schon}
Y. Makhlin, G. Sch\"{o}n, and A. Shnirman,
{\it Rev. Mod. Phys.} {\bf 73}, 357 (2001).


\bibitem{Pashkin}
Y.~A. Pashkin, T. Yamamoto, O. Astafiev, Y. Nakamura, D.~V. Averin, and J.~S. Tsai,
{\it Nature} {\bf 421}, 823 (2003).

\bibitem{Pashkin2}
T.~Yamamoto, Y.~A. Pashkin, O. Astafiev, Y. Nakamura, D.~V., and
J.~S. Tsai,
{\it Nature} {\bf 425}, 944 (2003).


\bibitem{martinis}
J.~M. Martinis, S. Nam, J. Aumentado, and C. Urbina,
{\it Phys. Rev. Lett.} {\bf 89}, 117901 (2002).

\bibitem{Smolin}
J. A. Smolin, and D. P. DiVincenzo,
{\it Phys. Rev. A} {\bf 53}, 2855 (1996).

\bibitem{poly}
J. C. Lagarias, J. A. Reeds, M. H. Wright, and P. E. Wright,
{\it SIAM J. Optim.} {\bf 9}, 112 (1998).

\bibitem{Gruska}
J. Gruska, \emph{Quantum Computing}, McGraw-Hill, New York (1999), p. 118.

\bibitem{Svore}
A. V. Aho and K. M. Svore, ``Compiling Quantum Circuits using the
Palindrome Transform", quant-ph/0311008.

\end{thebibliography}
\end{document}